\documentclass[article,nojss]{jss}

\usepackage{thumbpdf,lmodern}

\usepackage{framed,amsmath,amsfonts}
\usepackage{algorithm}
\usepackage{array}

\def\bzero{{\bf 0}}
\newcommand{\bbN}{\mathbb{N}}
\newcommand{\bbR}{\mathbb{R}}
\newcommand{\calA}{\mathcal{A}}

\newcommand{\calS}{\mathcal{S}}
\newcommand{\dlt}{\delta}

\author{J. Kenneth Tay \\ Stanford University
   \And Balasubramanian Narasimhan \\ Stanford University
   \And Trevor Hastie \\ Stanford University}
\Plainauthor{J. Kenneth Tay, Balasubramanian Narasimhan, Trevor Hastie}

\title{Elastic Net Regularization Paths for All Generalized Linear Models}
\Plaintitle{Elastic Net Paths for All GLMs}
\Shorttitle{Elastic Net Paths for All GLMs}

\Abstract{
  The lasso and elastic net are popular regularized regression models for supervised learning. \cite{Friedman2010} introduced a computationally efficient algorithm for computing the elastic net regularization path for ordinary least squares regression, logistic regression and multinomial logistic regression, while \cite{Simon2011} extended this work to Cox models for right-censored data. We further extend the reach of the elastic net-regularized regression to all generalized linear model families, Cox models with (start, stop] data and strata, and a simplified version of the relaxed lasso. We also discuss convenient utility functions for measuring the performance of these fitted models.
}

\Keywords{lasso, elastic net, $\ell_1$ penalty, regularization path, coordinate descent, generalized linear models, survival, Cox model}
\Plainkeywords{lasso, elastic net, l1 penalty, regularization path, coordinate descent, generalized linear models, survival, Cox model, relaxed lasso}

\Address{
  J. Kenneth Tay\\
  Department of Statistics\\
  Stanford University\\
  390 Jane Stanford Way\\
  Stanford, California 94305\\
  United States of America\\
  E-mail: \email{kjytay@stanford.edu}\\
  URL: \url{https://kjytay.github.io/} \\
  \\
  Balasubramanian Narasimhan\\
  Department of Biomedical Data Sciences, and\\
  Department of Statistics\\
  Stanford University\\
  390 Jane Stanford Way\\
  Stanford, CA 94305\\
  E-mail: \email{naras@stanford.edu}\\
  URL: \url{https://web.stanford.edu/~naras/} \\
  \\
  Trevor Hastie\\
  Department of Biomedical Data Sciences, and\\
  Department of Statistics\\
  Stanford University\\
  390 Jane Stanford Way\\
  Stanford, CA 94305\\
  E-mail: \email{hastie@stanford.edu}\\
  URL: \url{https://web.stanford.edu/~hastie/}
}

\begin{document}

\section[Introduction]{Introduction} \label{sec:intro}

Consider the standard supervised learning framework. We have data of the form $(x_1, y_1), \dots,$ $(x_n, y_n)$, where $y_i \in \bbR$ is the target and $x_i = (x_{i,1}, \dots, x_{i,p})^T \in \bbR^p$ is a vector of potential predictors. The ordinary least squares (OLS) model assumes that the response can be modeled as a linear combination of the covariates, i.e. $y_i = \beta_0 + x_i^T \beta$ for some coefficient vector $\beta \in \bbR^p$ and intercept $\beta_0 \in \bbR$. The parameters are estimated by minimizing the residual sum of squares (RSS):
\begin{equation}\label{eqn:ols}
    (\hat\beta_0,\hat\beta) = \underset{(\beta_0,\beta) \in \bbR^{p+1}}{\text{argmin}} \frac{1}{2n}\sum_{i=1}^n (y_i - \beta_0 - x_i^T \beta)^2.
\end{equation}
There has been a lot of research on regularization methods in the last two decades. We focus on the elastic net \citep{Zou2005} which minimizes the sum of the RSS and a regularization term which is a mixture of $\ell_1$ and $\ell_2$ penalties:
\begin{equation}\label{eqn:ols-enet}
    (\hat\beta_0,\hat\beta) = \underset{(\beta_0,\beta) \in \bbR^{p+1}}{\text{argmin}} \left[\frac{1}{2n}\sum_{i=1}^n (y_i - \beta_0 - x_i^T \beta)^2 + \lambda \left( \frac{1-\alpha}{2}\|\beta\|_2^2 + \alpha \|\beta\|_1 \right) \right].
\end{equation}
In the above, $\lambda \geq 0$ is a tuning parameter and $\alpha \in [0,1]$ is a higher level hyperparameter\footnote{If the square were removed from the $\ell_2$-norm penalty, it would be more natural to have $1-\alpha$ instead of $(1-\alpha)/2$ as its mixing parameter. The factor of $1/2$ compensates for the fact that a squared $\ell_2$-norm penalty is used, in the sense that the gradient of the penalty with respect to $\beta$ can be seen as a convex combination of the $\ell_1$ and $\ell_2$ penalty terms. We note also that there is a one-to-one correspondence between these two parameterizations for the penalty.}. We always fit a path of models in $\lambda$, but set a value of $\alpha$ depending on the type of prediction model we want. For example, if we want ridge regression \citep{Hoerl1970} we set $\alpha = 0$ and if we want the lasso \citep{Tibshirani1996} we set $\alpha = 1$. If we want a sparse model but are worried about correlations between features, we might set $\alpha$ close to but not equal to 1. The final value of $\lambda$ is usually chosen via cross-validation: we select the coefficients corresponding to the $\lambda$ value giving smallest cross-validated error as the final model.

The elastic net can be extended easily to generalized linear models (GLMs) \citep{Nelder1972} and Cox proportional hazards models \citep{Cox1972}. Instead of solving the minimization problem \eqref{eqn:ols-enet}, the RSS term in the objective function is replaced with a negative log-likelihood term or a negative log partial likelihood term respectively.

The \pkg{glmnet} \proglang{R} package \citep{Friedman2010} contains efficient functions for computing the elastic net solution for an entire path of values $\lambda_1 > \dots > \lambda_m$. The minimization problems are solved via cyclic coordinate descent \citep{vanderKooij2007}, with the core routines programmed in \proglang{FORTRAN} for computational efficiency. Earlier versions of the package contained specialized \proglang{FORTRAN} subroutines for a handful of popular GLMs and the Cox model for right-censored survival data. The package includes functions for performing $K$-fold cross-validation (CV), plotting coefficient paths and CV errors, and predicting on future data. The package can also accept the predictor matrix in sparse matrix format: this is especially useful in certain applications where the predictor matrix is both large and sparse. In particular, this means that we can fit unpenalized GLMs with sparse predictor matrices, something the \code{glm} function in the \pkg{stats} package cannot do.

From version 4.1 and later, \pkg{glmnet} is able to compute the elastic net regularization path for all GLMs, Cox models with (start, stop] data and strata, and a simplified version of the relaxed lasso. \citep{Hastie2020}.

\cite{Friedman2010} gives details on how the \pkg{glmnet} package computes the elastic net solution for ordinary least squares regression, logistic regression and multinomial logistic regression, while \cite{Simon2011} explains how the package fits regularized Cox models for right-censored data. This paper builds on these two earlier works. In Section~\ref{sec:glm}, we explain how the elastic net penalty can be applied to all GLMs and how we implement it in software. In Section~\ref{sec:cox}, we detail extensions to Cox models with (start, stop] data and strata. In Section~\ref{sec:relax}, we describe an implementation of the relaxed lasso implemented in the package, and in Section~\ref{sec:assess} we describe the package's functionality for assessing fitted models. We conclude with a summary and discussion.

\section{Regularized generalized linear models}\label{sec:glm}

\subsection{Overview of generalized linear models}
Generalized linear models (GLMs) \citep{Nelder1972} are a simple but powerful extension of OLS. A GLM consists of 3 parts:

\begin{itemize}
    \item A linear predictor: $\eta_i = x_i^T \beta$,
    \item A link function: $\eta_i = g(\mu_i)$, and
    \item A variance function as a function of the mean: $V = V(\mu_i)$.
\end{itemize}

The user gets to specify the link function $g$ and the variance function $V$. For one-dimensional exponential families, the family determines the variance function, which, along with the link, are sufficient to specify a GLM. More generally, modeling can proceed once the link and variance functions are specified via a quasi-likelihood approach (see \cite{McCullagh1983} for details); this is the approach taken by the quasi-binomial and quasi-Poisson models. The OLS model is a special case, with link $g(x) = x$ and constant variance function $V(\mu) = \sigma^2$ for some constant $\sigma^2$. More examples of GLMs are listed in Table \ref{tab:glm}.

\begin{table}[t!]
\centering
\begin{tabular}{p{1.2in} p{1.5in} p{2.8in}}
\hline
GLM family / Regression type & Response type & Representation in R \\ \hline
Gaussian & $\bbR$ & \code{gaussian()} \\
Logistic & $\{ 0, 1\}$ & \code{binomial()} \\
Probit & $\{ 0, 1\}$ & \code{binomial(link = "probit")} \\
Quasi-Binomial & $\{ 0, 1\}$ & \code{quasibinomial()} \\
Poisson & $\bbN_0 = \{ 0, 1, \dots \}$ & \code{poisson()} \\
Quasi-Poisson & $\bbN_0$ & \code{quasipoisson()} \\
Negative binomial & $\bbN_0$ & \code{MASS::negative.binomial(theta = 3)} \\
Gamma & $\bbR_+ = [0, \infty)$ & \code{Gamma()}  \\
Inverse Gaussian & $\bbR_+$ & \code{inverse.gaussian()}  \\
Tweedie & Depends on variance power parameter & \code{statmod::tweedie()} \\
\hline

\end{tabular}
\caption{\label{tab:glm} Examples of generalized linear models (GLMs) and their representations in \proglang{R}.}
\end{table}

The GLM parameter $\beta$ is determined by maximum likelihood estimation. Unlike OLS, there is no closed form solution for $\hat\beta$. Rather, it is typically computed via an iteratively reweighted least squares (IRLS) algorithm known as \textit{Fisher scoring}. In each iteration of the algorithm we make a quadratic approximation to the negative log-likelihood (NLL), reducing the minimization problem to a weighted least squares (WLS) problem. For GLMs with canonical link functions, the negative log-likelihood is convex in $\beta$, Fisher scoring is equivalent to the Newton-Raphson method and is guaranteed to converge to a global minimum. For GLMs with non-canonical links, the negative log-likelihood is not guaranteed to be convex\footnote{It is not true that the negative log-likelihood is always non-convex for non-canonical links. For example, it can be shown via direct computation that the negative log-likelihood for probit regression is convex in $\beta$.}. Also, Fisher scoring is no longer equivalent to the Newton-Raphson method and is only guaranteed to converge to a local minimum.

It is easy to fit GLMs in \proglang{R} using the \code{glm} function from the \pkg{stats} package; the user can specify the GLM to be fit using \code{family} objects. These objects capture details of the GLM such as the link function and the variance function. For example, the code below shows the \code{family} object associated with probit regression model:
\begin{CodeChunk}
\begin{CodeInput}
R> class(binomial(link = "probit"))
\end{CodeInput}
\begin{CodeOutput}
[1] "family"
\end{CodeOutput}
\begin{CodeInput}
R> str(binomial(link = "probit"))
\end{CodeInput}
\begin{CodeOutput}
List of 12
 $ family    : chr "binomial"
 $ link      : chr "probit"
 $ linkfun   : function (mu)  
 $ linkinv   : function (eta)  
 $ variance  : function (mu)  
 $ dev.resids: function (y, mu, wt)  
 $ aic       : function (y, n, mu, wt, dev)  
 $ mu.eta    : function (eta)  
 $ initialize: language ... # code to set up objects needed for the family
 $ validmu   : function (mu)  
 $ valideta  : function (eta)  
 $ simulate  : function (object, nsim)  
 - attr(*, "class")= chr "family"
\end{CodeOutput}
\end{CodeChunk}
The \code{linkfun}, \code{linkinv}, \code{variance} and \code{mu.eta} functions are used in fitting the GLM, and the \code{dev.resids} function is used in computing the deviance of the resulting model. By passing a class \code{"family"} object to the \code{family} argument of a \code{glm} call, \code{glm} has all the information it needs to fit the model. Here is an example of how one can fit a probit regression model in R:
\begin{CodeChunk}
\begin{CodeInput}
R> library(glmnet)
R> data(BinomialExample)
R> glm(y ~ x, family = binomial(link = "probit"))
\end{CodeInput}
\end{CodeChunk}
\subsection{Extending the elastic net to all GLM families}
To extend the elastic net to GLMs, we replace the RSS term in \eqref{eqn:ols-enet} with a negative log-likelihood term:
\begin{equation}\label{eqn:glm-enet}
    (\hat\beta_0,\hat\beta) = \underset{(\beta_0,\beta) \in \bbR^{p+1}}{\text{argmin}} \left[-\frac{1}{n}\sum_{i=1}^n \ell \left(y_i, \beta_0 + x_i^T \beta \right) + \lambda \left( \frac{1-\alpha}{2}\|\beta\|_2^2 + \alpha \|\beta\|_1 \right) \right],
\end{equation}
where $\ell \left(y_i, \beta_0 + x_j^T \beta \right)$ is the log-likelihood term associated with observation $i$. We can apply the same strategy as for GLMs to minimize this objective function. The key difference is that instead of solving a WLS problem in each iteration, we solve a penalized WLS problem.

The algorithm for solving \eqref{eqn:glm-enet} for a path of $\lambda$ values is described in Algorithm \ref{alg:glm-enet}. Note that in Step 2(a), we initialize the solution for $\lambda = \lambda_k$ at the solution obtained for $\lambda = \lambda_{k-1}$. This is known as a \textit{warm start}: since we expect the solution at these two $\lambda$ values to be similar, the algorithm will likely require fewer iterations than if we initialized the solution at zero.

\begin{algorithm}
\caption{ \em Fitting GLMs with elastic net penalty}
\label{alg:glm-enet}
\begin{enumerate}
\item Select a value of $\alpha \in [0, 1]$ and a sequence of $\lambda$ values $\lambda_1 > \ldots > \lambda_m$.

\item For $k = 1, \dots, m$:

\begin{enumerate}
    \item Initialize $(\hat\beta_0^{(0)}(\lambda_k), \hat\beta^{(0)}(\lambda_k)) = (\hat\beta_0(\lambda_{k-1}), \hat\beta(\lambda_{k-1}))$. For $k=1$, initialize $(\hat\beta_0^{(0)}(\lambda_k), \hat\beta^{(0)}(\lambda_k))= (0, \bzero)$. (Here, $(\hat\beta_0(\lambda_k), \hat\beta(\lambda_k))$ denotes the elastic net solution at $\lambda = \lambda_k$.)
    
    \item For $t = 0, 1, \ldots$ until convergence:
    
    \begin{enumerate}
        \item For $i = 1, \dots, n$, compute $\eta_i^{(t)} = \hat\beta_0^{(t)}(\lambda_k) + \hat\beta^{(t)}(\lambda_k)^T x_i$ and $\mu_i^{(t)} = g^{-1} \left( \eta_i^{(t)} \right)$.
    
        \item For $i = 1, \dots, n$, compute working responses and weights
        \begin{equation}
            z_i^{(t)} = \eta_i^{(t)} + \left( y_i - \mu_i^{(t)} \right) \big/ \dfrac{d\mu_i^{(t)}}{d\eta_i^{(t)}}, \quad w_i^{(t)} = \left(\dfrac{d\mu_i^{(t)}}{d\eta_i^{(t)}} \right)^2 \big/ V \left(\mu_i^{(t)} \right). \label{eqn:glm-irls-working}
        \end{equation}
        
        \item Solve the penalized WLS problem
        \begin{align}
            &(\hat\beta_0^{(t+1)}(\lambda_k),\hat\beta^{(t+1)}(\lambda_k)) \nonumber \\
            &= \underset{(\beta_0,\beta) \in \bbR^{p+1}}{\text{argmin}} \left[\frac{1}{2n}\sum_{i=1}^n w_i^{(t)} \left(z_i^{(t)} - \beta_0 - x_i^T \beta \right)^2 + \lambda_k \left( \frac{1-\alpha}{2}\|\beta\|_2^2 + \alpha \|\beta\|_1 \right) \right]. \label{eqn:wls-enet}
        \end{align}
    \end{enumerate}
\end{enumerate}
\end{enumerate}
\end{algorithm}

\subsection{Implementation details}
\begin{figure}[t!]
\centering
\includegraphics[width=\textwidth]{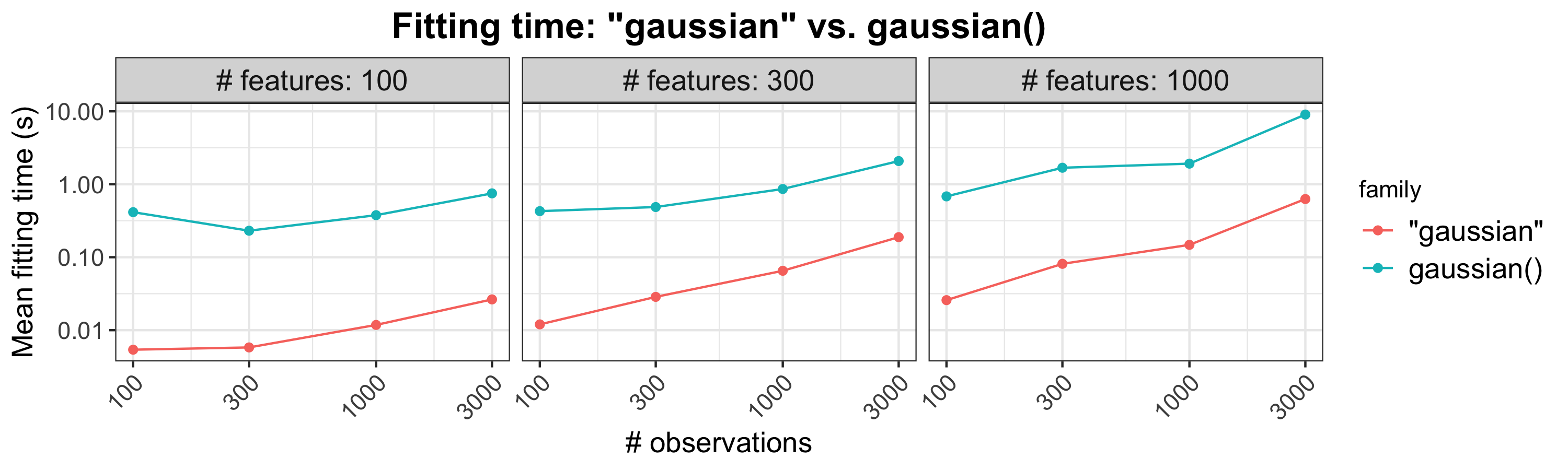}
\includegraphics[width=\textwidth]{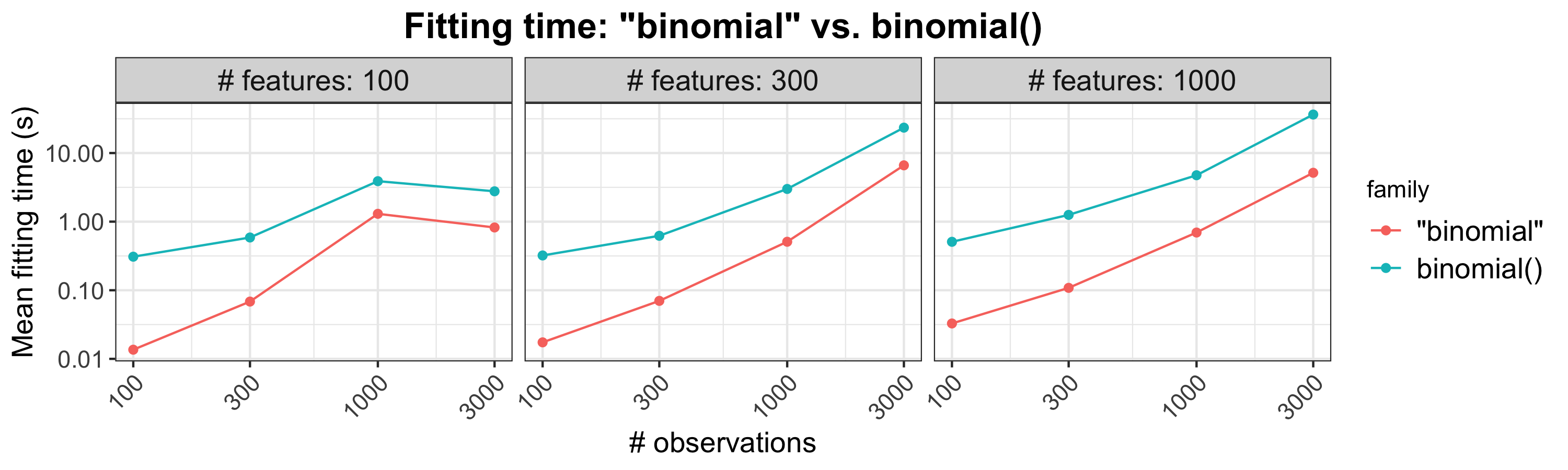}
\caption{\label{fig:glm-timing} The top plot compares model fitting times for \code{family = "gaussian"} and \code{family = gaussian()} for a range of problem sizes, while the plot below compares that for \code{family = "binomial"} and \code{family = binomial()}. Each point is the mean of 5 simulation runs. Note that both the $x$ and $y$ axes are on the log scale.}
\end{figure}

There are two main approaches we can take in implementing Algorithm \ref{alg:glm-enet}. In the original implementation of \pkg{glmnet}, the entire algorithm was implemented in \proglang{FORTRAN} for specific GLM families. In version 4.0 and later, we added a second implementation which implemented just the computational bottleneck, the penalized WLS problem in Step 2(b)iii, in \proglang{FORTRAN}, with the rest of the algorithm implemented in \proglang{R}. Here are the relative merits and disadvantages of the second approach compared to the first:
\begin{itemize}
    \item[\checkmark] Because the formulas for the working weights and responses in \eqref{eqn:glm-irls-working} are specific to each GLM, the first approach requires a new \proglang{FORTRAN} subroutine for each GLM family. This is tedious to manage, and also means that users cannot fit regularized models for their bespoke GLM families. The second approach allows the user to pass a class \code{"family"} object to \code{glmnet}: the working weights and responses can then be computed in \proglang{R} before the \proglang{FORTRAN} subroutine solves the resulting penalized WLS problem.
    
    \item[\checkmark] As written, Algorithm \ref{alg:glm-enet} is a proximal Newton algorithm with a constant step size of 1, and hence it may not converge in certain cases. To ensure convergence, we can implement step-size halving after Step 2(b)iii: as long as the objective function \eqref{eqn:glm-enet} is not decreasing, set $\hat\beta^{(t+1)}(\lambda_k) \leftarrow \hat\beta^{(t)}(\lambda_k) + \frac{1}{2}\left[ \hat\beta^{(t+1)}(\lambda_k) - \hat\beta^{(t)}(\lambda_k)\right]$ (with a similar formula for the intercept). Since the objective function involves a log-likelihood term, the formula for the objective function differs across GLMs, and the first approach has to maintain different subroutines for step-size halving. For the second approach, we can write a single function that takes in the class \code{"family"} object (along with other necessary parameters) and returns the objective function value.
    
    \item[$\times$] It is computationally less efficient than the first approach because (i) \proglang{R} is generally slower than \proglang{FORTRAN}, and (ii) there is overhead associated with constant switching between \proglang{R} and \proglang{FORTRAN}. Some timing comparisons for Gaussian and logistic regression with the default parameters are presented in Figure \ref{fig:glm-timing}. The second approach is 10 to 15 times as slow than the first approach.
    
    \item[$\times$] Since each GLM family has its own set of \proglang{FORTRAN} subroutines in the first approach, it allows for special computational tricks to be employed in each situation. For example, with \code{family = "gaussian"}, the predictors can be centered once upfront to have zero mean and Algorithm \ref{alg:glm-enet} can be run ignoring the intercept term.
\end{itemize}

We stress that both approaches have been implemented in \pkg{glmnet}. Users should use the first implementation for the most popular GLM families including OLS (Gaussian regression), logistic regression and Poisson regression (see \code{glmnet}'s documentation for the full list of such families), and use the second implementation for all other GLM families. For example, the code below shows two equivalent ways to fit a regularized Poisson regression model:
\begin{CodeChunk}
\begin{CodeInput}
R> data(PoissonExample)
R> glmnet(x, y, family = "poisson")
R> glmnet(x, y, family = poisson())
\end{CodeInput}
\end{CodeChunk}
The first call specifies the GLM family as a character string to the \code{family} argument, invoking the first implementation. The second call passes a class \code{"family"} object to the \code{family} argument instead of a character string, invoking the second implementation. One would never run the second call in practice though, as it returns the same result as the first call but takes longer to fit. The example below fits a regularized quasi-Poisson model that allows for overdispersion, a family that is only available via the second approach:
\begin{CodeChunk}
\begin{CodeInput}
R> glmnet(x, y, family = quasipoisson())
\end{CodeInput}
\end{CodeChunk}
\subsection{Details on the penalized WLS subroutine}\label{sec:wls}

Since the penalized WLS problem in Step 2(b)iii of Algorithm \ref{alg:glm-enet} is the computational bottleneck, we elected to implement it in \proglang{FORTRAN}. Concretely, the subroutine solves the problem
\begin{align}
\underset{(\beta_0,\beta) \in \bbR^{p+1}}{\text{minimize}} \qquad& \frac{1}{2n}\sum_{i=1}^n w_i \left(z_i - \beta_0 - x_i^T \beta \right)^2 + \lambda_k \sum_{j=1}^p \gamma_j \left( \frac{1-\alpha}{2}\beta_j^2 + \alpha |\beta_j | \right) \\
\text{subject to} \qquad& L_j \leq \beta_j \leq U_j, \quad j = 1, \dots, p. \label{eqn:wls-enet2}
\end{align}

This is the same problem as \eqref{eqn:wls-enet} except for two things. First, the penalty placed on each coefficient $\beta_j$ has its own multiplicative factor $\gamma_j$. (\eqref{eqn:wls-enet2} reduces to \eqref{eqn:wls-enet} if $\gamma_j = 1$ for all $j$, which is the default value for the \code{glmnet} function.) This allows the user to place different penalty weights on the coefficients. An instance where this is especially useful is when the user always wants to include feature $j$ in the model: in that case the user could set $\gamma_j = 0$ so that $\beta_j$ is unpenalized. Second, the coefficient $\beta_j$ is constrained to lie in the interval $[L_j, U_j]$. (\code{glmnet}'s default is $L_j = -\infty$ and $U_j = \infty$ for all $j$, i.e. no constraints on the coefficients.) One example where these constraints are useful is when we want a certain $\beta_j$ to always be non-negative or always non-positive.

The \proglang{FORTRAN} subroutine solves \eqref{eqn:wls-enet2} by cyclic coordinate descent: see \cite{Friedman2010} for details. Here we describe one major computational trick that was not covered in that paper: the application of \textit{strong rules} \citep{Tibshirani2012}.

In each iteration of cyclic coordinate descent, the solver has to loop through all $p$ features to update the corresponding model coefficients. This can be time-consuming if $p$ is large, and is potentially wasteful if the solution is sparse: most of the $\beta_j$ would remain at zero. If we know a priori which predictors will be ``active'' at the solution (i.e. have $\beta_j \neq 0$), we could perform cyclic coordinate descent on just those coefficients and leave the others untouched. The set of ``active" predictors is known as the \textit{active set}. Strong rules are a simple yet powerful heuristic for guessing what the active set is, and can be combined with the Karush-Kuhn-Tucker (KKT) conditions to ensure that we get the exact solution. (The set of predictors determined by the strong rules is known as the \textit{strong set}.) We describe the use of strong rules in solving \eqref{eqn:wls-enet2} fully in Algorithm \ref{alg:wls-enet2}.

\begin{algorithm}
\caption{ \em Solving penalized WLS \eqref{eqn:wls-enet2} with strong rules}
\label{alg:wls-enet2}

Assume that we are trying to solve for $\hat\beta(\lambda_k)$ for some $k = 1, \dots, m$, and that we have already computed $\hat\beta(\lambda_{k-1})$. (If $k = 1$, set $\hat\beta(\lambda_{k-1}) = \bzero$.)

\begin{enumerate}
\item Initialize the strong set $\calS_{\lambda_k} = \{ j: \hat\beta(\lambda_{k-1})_j \neq 0 \}$.

\item Check the strong rules: for $j = 1, \dots, p$, include $j$ in $\calS_{\lambda_k}$ if
\begin{equation*}
    \left| x_j^T \left\{ y - X \hat\beta(\lambda_{k-1}) \right\} \right| > \alpha \left[ \lambda_k - (\lambda_{k-1} - \lambda_k) \right] \gamma_j.
\end{equation*}

\item Perform cyclic coordinate descent only for features in $\calS_{\lambda_k}$.

\item Check that the KKT conditions hold for each $j = 1, \dots, p$. If the conditions hold for all $j$, we have the exact solution. If the conditions do not hold for some features, include them in the strong set $\calS_{\lambda_k}$ and go back to Step 3.
\end{enumerate}
\end{algorithm}

Finally, we note that in some applications, the design matrix $X$ is sparse. In these settings, computational savings can be reaped by representing $X$ in a sparse matrix format and performing matrix manipulations with this form. To leverage this property of the data, we have a separate \proglang{FORTRAN} subroutine that solves \eqref{eqn:wls-enet2} when $X$ is in sparse matrix format.

\subsection{Other useful functionality}

In this section, we mention other use functionality that the \pkg{glmnet} package provides for fitting elastic net models.

For fixed $\alpha$, \code{glmnet} solves \eqref{eqn:glm-enet} for a path of $\lambda$ values. While the user has the option of specifying this path of values using the \code{lambda} option, it is recommended that the user let \code{glmnet} compute the sequence on its own. \code{glmnet} uses the arguments passed to it to determine the value of $\lambda_{max}$, defined to be the smallest value of $\lambda$ such that the estimated coefficients would be all equal to zero\footnote{We note that when $\alpha = 0$, $\lambda_{max}$ is infinite, i.e. all coefficients will always be non-zero for finite $\lambda$. To avoid such extreme values of $\lambda_{max}$, if $\alpha < 0.001$ we return the $\lambda_{max}$ value for $\alpha = 0.001$.}. The program then computes $\lambda_{min}$ such that the ratio $\lambda_{min} / \lambda_{max}$ is equal to \code{lambda.min.ratio} (default $10^{-2}$ if the number of variables exceeds the number of observations, $10^{-4}$ otherwise). Model \eqref{eqn:glm-enet} is then fit for \code{nlambda} $\lambda$ values (default 100) starting at $\lambda_{max}$ and ending at $\lambda_{min}$ which are equally spaced on the log scale.

\begin{figure}[t!]
\centering
\includegraphics{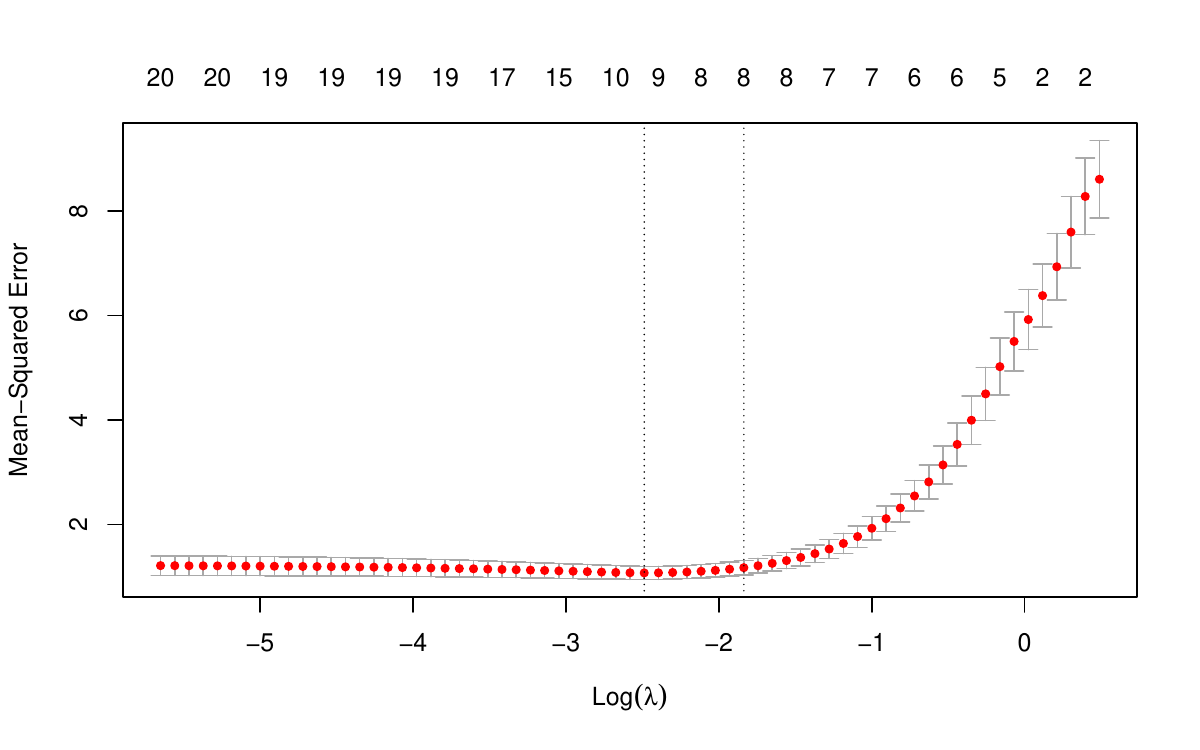}
\caption{\label{fig:cv-curve} Example output for plotting a \code{cv.glmnet} object: a plot of CV error against $\log(\lambda)$. The error bars correspond to $\pm 1$ standard error. The left vertical line corresponds to the minimum error while the right vertical line corresponds to the largest value of $\lambda$ such that the CV error is within one standard error of the minimum. The top of the plot is annotated with the size of the models, i.e. the number of predictors with non-zero coefficient.}
\end{figure}

In practice, it common to choose the value of $\lambda$ via cross-validation (CV). The \code{cv.glmnet} function is a convenience function that runs CV for the $\lambda$ tuning parameter. The returned object has class \code{"cv.glmnet"}, which comes equipped with \code{plot}, \code{coef} and \code{predict} methods. The \code{plot} method produces a plot of CV error against $\lambda$ (see Figure \ref{fig:cv-curve} for an example.) As mentioned earlier, we prefer to think of $\alpha$ as a higher level hyperparameter whose value depends on the type of prediction model we want. Nevertheless, the code below shows how the user can perform CV for $\alpha$ manually using a for loop. Care must be taken to ensure that the same CV folds are used across runs for the CV errors to be comparable.
\begin{CodeChunk}
\begin{CodeInput}
R> alphas <- c(1, 0.8, 0.5, 0.2, 0)
R> fits <- list()
R> fits[[1]] <- cv.glmnet(x, y, keep = TRUE)
R> foldid <- fits[[1]]$foldid
R> for (i in 2:length(alphas)) {
+      fits[[i]] <- cv.glmnet(x, y, alpha = alphas[i], foldid = foldid)
+  }
\end{CodeInput}
\end{CodeChunk}

The returned \code{cv.glmnet} object contains estimated standard errors for the model CV error at each $\lambda$ value. (We note that the method for obtaining this estimates is crude and are generally too small due to correlations across CV folds.) By default, the \code{predict} method returns predictions for the model at the \code{"lambda.1se"} value, i.e. the value of $\lambda$ that gives the most regularized model such that the CV error is within one standard error of the minimum. To get predictions at the $\lambda$ value which gives the minimum CV error, the \code{s = "lambda.min"} argument is passed.
\begin{CodeChunk}
\begin{CodeInput}
R> cfit <- cv.glmnet(x, y)
R> predict(cfit, x)
R> predict(cfit, x, s = "lambda.min")
\end{CodeInput}
\end{CodeChunk}
In large data settings, it may take some time to fit the entire sequence of elastic net models. \code{glmnet} and \code{cv.glmnet} come equipped with a progress bar which can be displayed with the argument \code{trace.it = TRUE}. This gives the user a sense of how model fitting is progressing.

The \pkg{glmnet} package provides a convenience function \code{bigGlm} for fitting a single \textit{unpenalized} GLM but allowing all the options of \code{glmnet}. In particular, the user can set upper and/or lower bounds on the coefficients, and can provide the \code{x} matrix in sparse matrix format: options that are not available for the \code{stats::glm} function.
\begin{CodeChunk}
\begin{CodeInput}
R> data(BinomialExample)
R> fit <- bigGlm(x, y, family = "binomial", lower.limits = -1)
\end{CodeInput}
\end{CodeChunk}
\section{Regularized Cox proportional hazards models}\label{sec:cox}

We assume the usual survival-analysis framework. Instead of having $y_i \in \bbR$ as a response, we have instead $(y_i, \dlt_i) \in \bbR_+ \times \{ 0, 1\}$. Here $y_i$ is the observed time for observation $i$, and $\dlt_i = 1$ if $y_i$ is the failure time and $\dlt_i = 0$ if it is the right-censoring time. The Cox proportional hazards model \citep{Cox1972} is a commonly used model for the relationship between the predictor variables and survival time. It assumes a semi-parametric form for the hazard function
\begin{equation*}
    h_i(t) = h(t) e^{x_i^T \beta},
\end{equation*}
where $h_i(t)$ is the hazard for observation $i$ at time $t$, $h$ is the baseline hazard for the entire population of observations, and $\beta \in \bbR^p$ is the vector of coefficients to be estimated. Let $t_1 < \dots < t_m$ denote the unique failure times and let $j(i)$ denote the index of the observation failing at time $t_i$. (Assume for the moment that the $y_i$'s are unique.) If $y_j \geq t_i$, we say that observation $j$ is \textit{at risk} at time $t_i$. Let $R_i$ denote the \textit{risk set} at time $t_i$. $\beta$ is estimated by maximizing the partial likelihood
\begin{equation}\label{eqn:pl}
    L(\beta) = \prod_{i=1}^m \frac{ e^{x_{j(i)}^T \beta}}{\sum_{j \in R_i} e^{x_j^T \beta}}.
\end{equation}
It is the conditional likelihood that the failure occurs for observation $j(i)$ given all the observations at risk. Maximizing the partial likelihood is equivalent to minimizing the negative log partial likelihood
\begin{equation}\label{eqn:lpl}
    -\ell(\beta) = \frac{2}{n}\sum_{i=1}^m \left[ -x_{j(i)}^T \beta + \log \left( \sum_{j \in R_i} e^{x_j^T \beta} \right) \right].
\end{equation}
We put a negative sign in front of $\ell$ so that $\ell$ denotes the log partial likelihood, and the scale factor $2/n$ is included for convenience. Note also that the model does not have an intercept term $\beta_0$, as it cancels out in the partial likelihood. \cite{Simon2011} proposed an elastic-net regularization path version for the Cox model, as well as Algorithm \ref{alg:cox-enet} for solving the minimization problem.

\begin{algorithm}
\caption{\em Fitting Cox models with elastic net penalty}
\label{alg:cox-enet}
\begin{enumerate}
\item Select a value of $\alpha \in [0, 1]$ and a sequence of $\lambda$ values $\lambda_1 > \ldots > \lambda_m$. Define $\hat\beta(\lambda_0) = \bzero$.

\item For $\ell = 1, \dots, m$:

\begin{enumerate}
    \item Initialize $\hat\beta(\lambda_\ell) = \hat\beta(\lambda_{\ell-1})$.
    
    \item For $t = 0, 1, \ldots$ until convergence (outer loop):
    
    \begin{enumerate}
        \item For $k = 1, \dots, n$, compute $\eta_k^{(t)} = \hat\beta(\lambda_\ell)^T x_k$.
    
        \item For $k = 1, \dots, n$, compute
        \begin{align}
        \ell' \left(\eta^{(t)} \right)_k &= \dlt_k - e^{\eta_k^{(t)}} \sum_{i \in C_k} \left( \frac{1}{\sum_{j \in R_i} e^{\eta_j^{(t)}} } \right), \label{eqn:lpl-grad}   \\
        \ell'' \left(\eta^{(t)} \right)_{k,k} &=
        \sum_{i \in C_k} \left[ \frac{e^{\eta_k^{(t)}} \sum_{j \in R_i} e^{\eta_j^{(t)}} - \left( e^{\eta_k^{(t)}} \right)^2 }{ \left( \sum_{j \in R_i} e^{\eta_j^{(t)}} \right)^2 } \right], \label{eqn:lpl-hess} \\
        w_k^{(t)} &= -\ell'' \left(\eta^{(t)} \right)_{k,k}, \label{eqn:lpl-weights} \\
        z_k^{(t)} &= \eta_k^{(t)} + \frac{\ell' \left(\eta^{(t)} \right)_k}{\ell'' \left(\eta^{(t)} \right)_{k,k}}, \label{eqn:lpl-response}
        \end{align}
        where $C_k$ is the set of failure times $i$ such that $t_i < y_k$ (i.e. times for which observation $k$ is still at risk.)
        
        \item Solve the penalized WLS problem (inner loop):
        \begin{equation*}
            \hat\beta(\lambda_\ell) = \underset{\beta \in \bbR^p}{\text{argmin}} \left[\frac{1}{2}\sum_{k=1}^n w_k^{(t)} \left(z_k^{(t)} - x_k^T \beta \right)^2 + \lambda_\ell \left( \frac{1-\alpha}{2}\|\beta\|_2^2 + \alpha \|\beta\|_1 \right) \right].
        \end{equation*}
    \end{enumerate}
\end{enumerate}
\end{enumerate}
\end{algorithm}

Algorithm \ref{alg:cox-enet} has the same structure as Algorithm \ref{alg:glm-enet} except for different formulas for computing the working responses and weights. (We note that these formulas implicitly approximate the Hessian of the log partial likelihood by a diagonal matrix with the Hessian's diagonal entries.) This means that we can leverage the fast implementation of the penalized WLS problem in Section \ref{sec:wls} for an efficient implementation of Algorithm \ref{alg:cox-enet}. (As a small benefit, it also means that we can fit regularized Cox models when the design matrix $X$ is sparse.) Such a model can be fit with \pkg{glmnet} by specifying \code{family = "cox"}. The response provided needs to be a \code{Surv} object from the \pkg{survival} package \citep{survival-package}.
\begin{CodeChunk}
\begin{CodeInput}
R> glmnet(x, y, family = "cox")
\end{CodeInput}
\end{CodeChunk}
The computation of these $w_k$'s and $z_k$'s can be a computational bottleneck if not implemented carefully: since the $C_k$ and $R_i$ have $O(n)$ elements, a naive implementation takes $O(n^2)$ time. \cite{Simon2011} exploit the fact that, once the observations are sorted in order of the observed times $y_i$, the risk sets are nested ($R_{i+1} \subseteq R_i$ for all $i$) and the $w_k$'s and $z_k$'s can be computed in $O(n)$ time.

If our data contains tied observed times, \code{glmnet} uses the Breslow approximation of the partial likelihood for ties \citep{breslow1972} and maximizes the elastic net-regularized version of this approximation instead. See \cite{Simon2011} for details.

\subsection{Extending regularized Cox models to (start, stop] data}

Instead of working with right-censored responses, the Cox model can be extended to work with responses which are a pair of times (called the ``start time'' and ``stop time''), with the possibility of the stop time being censored. This is an instantiation of the counting process framework proposed by \cite{Andersen1982}, and the right-censored data set-up is a special case with the start times all being equal to zero.

As noted in \cite{Therneau2000}, (start, stop] responses greatly increase the flexibility of the Cox model, allowing for
\begin{itemize}
    \item Time-dependent covariates,
    \item Time-dependent strata,
    \item Left truncation,
    \item Multiple time scales,
    \item Multiple events per subject,
    \item Independent increment, marginal, and conditional models for correlated data, and
    \item Various forms of case-cohort models.
\end{itemize}

From a data analysis viewpoint, this extension amounts to requiring just one more variable: the \code{time} variable is replaced by \code{(start, stop]} variables, with \code{(start, stop]} indicating the interval where the unit is at risk. The \pkg{survival} package provides the function \code{tmerge} to aid in the creation of such datasets.

For this more general setup, inference for $\beta$ can proceed as before. The formulas for the partial likelihood and negative log partial likelihood (Equations \eqref{eqn:pl} and \eqref{eqn:lpl}) remain the same; what changes is the definition of what it means for an observation to be at risk at time $t_i$. If we let $(y_{1j}, y_{2j}]$ denote the (start, stop] times for observation $j$, then observation $j$ is at risk at time $t_i$ if and only if $t_i \in (y_{1j}, y_{2j}]$. Similarly, the elastic net-regularized version of the Cox model for (start, stop] data can be fitted using Algorithm \ref{alg:cox-enet} with this new definition of what it means for an observation to be at risk at a failure time.

With (start, stop] data, it is no longer true that the risk sets are nested. For example, if $t_i < y_{1j} < t_{i+1} < y_{2j}$, then $j \in R_{i+1}$ but $j \notin R_i$. However, as Algorithm \ref{alg:cox-update} shows, it is still possible to compute the working responses and weights in $O(n \log n)$ time. In fact, only the ordering of observations (Step 1) requires $O(n \log n)$ time: the rest of the algorithm requires just $O(n)$ time. Since the ordering of observations never changes, the results of Step 1 can be cached, meaning that only the first run of Algorithm \ref{alg:cox-update} requires $O(n \log n)$ time, and future runs just need $O(n)$ time.

\begin{algorithm}
\caption{\em Computing working responses and weights for Algorithm \ref{alg:cox-enet}}
\label{alg:cox-update}
Input: $\eta_j = x_j^T \hat\beta$ (where $\hat\beta$ is the current estimate for $\beta$), $(y_{1j}, y_{2j}]$, $\dlt_j$ for $j = 1, \dots, n$. For simplicity, assume that the observations are ordered by ascending stop time, i.e. $y_{21} < \dots < y_{2n}$. As before, let $t_1 < \dots, t_m$ denote the failure times in increasing order.
\begin{enumerate}
\item Get the ordering for the observations according to start times. Let $start(j)$ denote the index for the observation with the $j$th earliest start time.

\item Compute the risk set sums $RSS_i = \sum_{j \in R_i} e^{\eta_j}$, $i = 1, \dots, m$ using the following steps:
\begin{enumerate}
    \item For $j = 1, \dots, n$, set $RSS_j \leftarrow \sum_{\ell = j}^n e^{\eta_{\ell}}$.
    \item Set $curr \leftarrow 0$, $i \leftarrow m$, $start\_idx \leftarrow n$.
    \item While $i > 0$ and $start\_idx > 0$:
    \begin{enumerate}
        \item If $y_{1 start(start\_idx)} < t_i$, set $RSS_{j(i)} \leftarrow RSS_{j(i)} - curr$ and $i \leftarrow i - 1$.
        
        \item If not, set $curr \leftarrow curr + e^{\eta_{start(start\_idx)}}$ and $start\_idx \leftarrow start\_idx - 1$.
    \end{enumerate}
    \item Take just the elements of $RSS$ corresponding to death times, i.e. set $RSS_i \leftarrow RSS_{j(i)}$.
\end{enumerate}

\item Compute the partial sums $RSK_k = \sum_{i \in C_k} \frac{1}{RSS_i}$, $k = 1, \dots, n$ using the following steps:
\begin{enumerate}
    \item For $i = 1, \dots, m$, set $RD_i \leftarrow \sum_{\ell = 1}^i \frac{1}{RSS_i}$. Set $RD_0 \leftarrow 0$.
    \item For $k = 1, \dots, n$, set $D_k \leftarrow \sum_{\ell = 1}^k \dlt_\ell$.
    \item For $k = 1, \dots, n$, set $RSK_k \leftarrow RD_{D_k}$.
    \item Set $curr \leftarrow 0$, $i \leftarrow 1$, $start\_idx \leftarrow 1$.
    \item While $i \leq m$ and $start\_idx \leq n$:
    \begin{enumerate}
        \item If $y_{1 start(start\_idx)} < t_i$, set $RSK_{start(start\_idx)} \leftarrow RSK_{start(start\_idx)} - curr$ and $start\_idx \leftarrow start\_idx + 1$.
        \item If not, set $curr \leftarrow curr + \frac{1}{RSS_i}$ and $i \leftarrow i + 1$.
    \end{enumerate}
\end{enumerate}

\item Compute the partial sums $RSKSQ_k = \sum_{i \in C_k} \frac{1}{RSS_i^2}$, $k = 1, \dots, n$ in a similar manner as Step 3.

\item Compute $\ell'(\eta)_k$ amd $\ell''(\eta)_{k,k}$ using the formulas \eqref{eqn:lpl-grad} and \eqref{eqn:lpl-hess}:
\begin{equation*}
    \ell'(\eta)_k = \dlt_k - e^{\eta_k} \cdot RSK_k, \qquad \ell''(\eta)_{k,k} = \left( e^{\eta_k} \right)^2 \cdot RSKSQ_k - e^{\eta_k} \cdot RSK_k.
\end{equation*}

\item Compute the working responses and weights using the formulas \eqref{eqn:lpl-weights} and \eqref{eqn:lpl-response}.

\end{enumerate}
\end{algorithm}

The differences between right-censored data and (start, stop] data for Cox models are hidden from the user, in that the function call for (start, stop] data is exactly the same as that for right-censored data. The difference is in the type of \code{Surv} object that is passed for the response \code{y}. \code{glmnet} checks for the \code{Surv} object type before routing to the correct internal subroutine.

\subsection{Stratified Cox models}

An extension of the Cox model is to allow for strata. These strata divide the units into disjoint groups, with each group having its own baseline hazard function but having the same values of $\beta$. Specifically, if the units are divided into $K$ strata, then the stratified Cox model assumes that a unit in stratum $k$ has the hazard function
\begin{equation*}
    h_i(t) = h_k(t) e^{x_i^T \beta},
\end{equation*}
where $h_k(t)$ is the shared baseline hazard for all units in stratum $k$. In several applications, allowing different subgroups to have different baseline hazards approximates reality more closely. For example, it might be reasonable to have different baseline hazards based on gender in clinical trials, or a separate baseline for each center in multi-center trials.

In this setting, the negative log partial likelihood is
\begin{equation*}
    \ell(\beta) = \sum_{k=1}^K \ell_k(\beta),
\end{equation*}
where $\ell_k(\beta)$ is exactly \eqref{eqn:lpl} but considering just the units in stratum $k$. Since the negative log partial likelihood decouples across strata (conditional on $\beta$), regularized versions of stratified Cox models can be fit using a slightly modified version of Algorithm \ref{alg:cox-enet}.

To fit an unpenalized stratified Cox model, the \pkg{survival} package has a special \code{strata} function that allows users to specify the strata variable in formula syntax. Since \code{glmnet} does not work with formulas, we needed a different approach for specifying strata. To fit regularized stratified Cox models in \pkg{glmnet}, the user needs to add a \code{strata} attribute to the response \code{y}. \code{glmnet} checks for the presence of this attribute and if it is present, it fits a stratified Cox model. We note that the user cannot simply add the attribute manually because \proglang{R} drops attributes when subsetting vectors. Instead, the user should use the \code{stratifySurv} function to add the \code{strata} attribute. (\code{stratifySurv} creates an object of class \code{"stratifySurv"} that inherits from the class \code{"Surv"}, ensuring that \pkg{glmnet} can reassign the \code{strata} attribute correctly after any subsetting.) The code below shows an example of how to fit a regularized stratified Cox model with \pkg{glmnet}; there are a total of 20 observations, with the first 10 belonging to the first strata and the rest belonging to the second strata.
\begin{CodeChunk}
\begin{CodeInput}
R> strata <- c(rep(1, 10), rep(2, 10))
R> y2 <- stratifySurv(y, strata)
R> glmnet(x, y2, family = "cox")
\end{CodeInput}
\end{CodeChunk}
\subsection{Plotting survival curves}

The beauty of the Cox partial likelihood is that the baseline hazard, $h_0(t)$, is not required for inference on the model coefficients $\beta$. However, the estimated hazard is often of interest to users. The \pkg{survival} package already has a well-established \code{survfit} method that can produce estimated survival curves from a fitted Cox model. \pkg{glmnet} implements a \code{survfit} method for regularized Cox models fit by \code{glmnet} by creating the \code{coxph} object corresponding to the model and calling \code{survival::survfit}.

\begin{figure}[t!]
\centering
\includegraphics[width=0.5\textwidth]{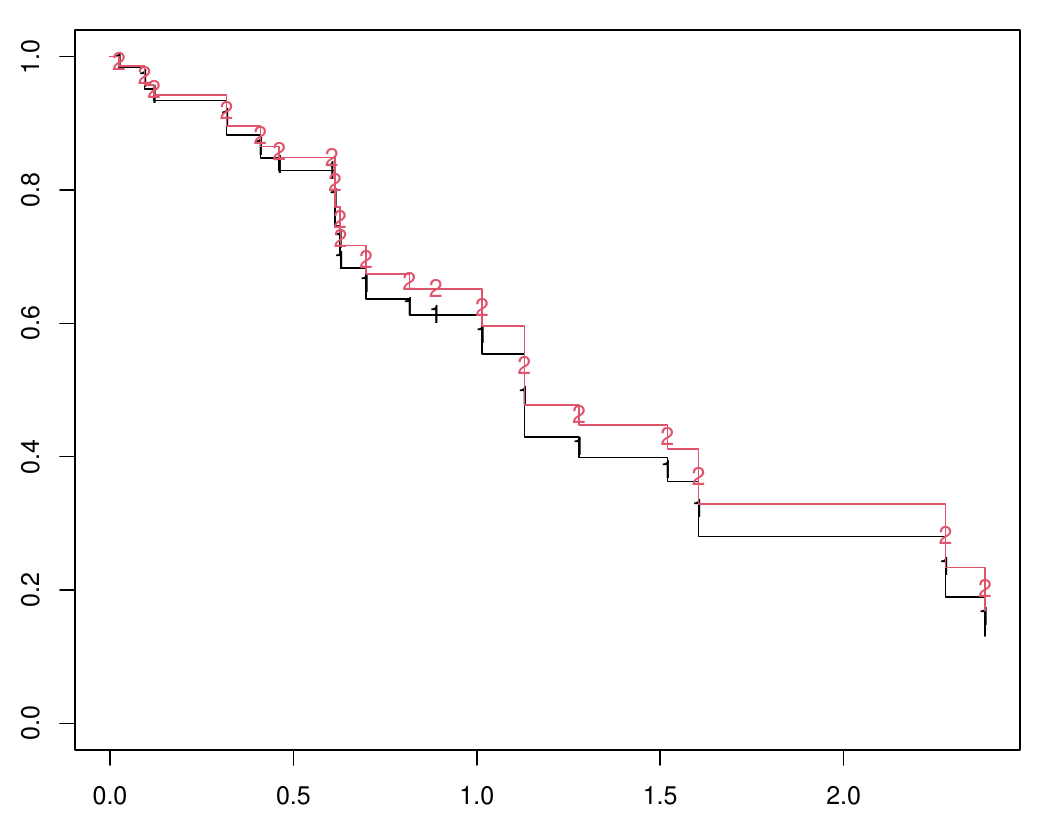}
\caption{\label{fig:survfit} An illustration of the plotted \code{survfit} object. One survival curve is plotted for each individual represented in the \code{newx} argument.}
\end{figure}

The code below is an example of calling \code{survfit} for \code{coxnet} objects for a particular value of the $\lambda$ tuning parameter (in this case, $\lambda = 0.05$). Note that we had to pass the original design matrix \code{x} and response \code{y} to the \code{survfit} call: they are needed for \code{survfit.coxnet} to reconstruct the required \code{coxph} object. The survival curves are computed for the individuals represented in \code{newx}: we get one curve per individual, as seen in Figure \ref{fig:survfit}.
\begin{CodeChunk}
\begin{CodeInput}
R> fit <- glmnet(x, y, family = "cox")
R> sf_obj <- survfit(fit, s = 0.05, x = x, y = y, newx = x[1:2, ])
R> plot(sf_obj, col = 1:2, mark.time = TRUE, pch = "12")
\end{CodeInput}
\end{CodeChunk}

The \code{survfit} method is available for Cox models fitted by \code{cv.glmnet} as well. By default, the survival curves are computed for the \code{lambda.1se} value of the $\lambda$ hyperparameter. The user can use the code below to compute the survival curve at the \code{lambda.min} value:
\begin{CodeChunk}
\begin{CodeInput}
R> c.fit <- cv.glmnet(x, y, family = "cox", nfolds = 5)
R> survfit(cfit, s = "lambda.min", x = x, y = y, newx = x[1:2, ])
\end{CodeInput}
\end{CodeChunk}
\section{The relaxed lasso}\label{sec:relax}

Due to the regularization penalty, the lasso tends to shrink the coefficient vector $\hat\beta$ toward zero. The relaxed lasso \citep{Meinshausen2007} was introduced as a way to undo the shrinkage inherent in the lasso estimator. Through extensive simulations, \cite{Hastie2020} conclude that the relaxed lasso performs well in terms of predictive performance across a range of scenarios. It was found to perform just as well as the lasso in low signal-to-noise (SNR) scenarios and nearly as well as best subset selection in high (SNR) scenarios. It also has a considerable advantage over best subset and forward stepwise regression when the number of variables, $p$, is large. In this section, we describe the simplified version of the relaxed lasso proposed by \cite{Hastie2020} and give details on how it is implemented in \pkg{glmnet}.

For simplicity, we describe the method for the OLS setting (\code{family = "gaussian"}) and for the lasso ($\alpha = 1$). For a given tuning parameter $\lambda$, let $\hat\beta^{lasso}(\lambda) \in \bbR^p$ denote the lasso estimator for this value of $\lambda$. Let $\calA_\lambda$ denote the active set of the lasso estimator, and let $\hat\beta^{LS}_{A_\lambda} \in \bbR^{|\calA_\lambda|}$ denote the OLS coefficients obtained by regressing $y$ on $X_{\calA_\lambda}$ (i.e. the subset of columns of $X$ which correspond to features in the active set $\calA_\lambda$). Let $\hat\beta^{LS}(\lambda) \in \bbR^p$ denote the OLS coefficients $\hat\beta^{LS}_{A_\lambda}$ padded with zeros to match the zeros of the lasso solution. The (simplified version of the) relaxed lasso estimator is given by
\begin{equation}\label{eqn:relaxed}
    \hat\beta^{relax}(\lambda, \gamma) = \gamma \hat\beta^{lasso}(\lambda) + (1 - \gamma) \hat\beta^{LS}(\lambda),
\end{equation}
where $\gamma \in [0, 1]$ is a hyperparameter, similar to $\alpha$. In other words, the relaxed lasso estimator is a convex combination of the lasso estimator and the OLS estimator for the lasso's active set.

The relaxed lasso can be fit with \code{glmnet} function in \pkg{glmnet} by setting the argument \code{relax = TRUE}:
\begin{CodeChunk}
\begin{CodeInput}
R> data(QuickStartExample)
R> fit <- glmnet(x, y, relax = TRUE)
\end{CodeInput}
\end{CodeChunk}
When called with this option, \code{glmnet} first runs the lasso (Algorithm \ref{alg:glm-enet} with $\alpha = 1$) to obtain the lasso estimates $\hat\beta^{lasso}(\lambda_k)$ and the active sets $\calA_{\lambda_k}$ for a path of hyperparameter values $\lambda_1 > \dots > \lambda_m$. It then goes down this sequence of hyperparameter values again, fitting the unpenalized model of $y$ on each $X_{\calA_{\lambda_k}}$ to obtain $\hat\beta^{LS}(\lambda_k)$. The refitting is done in an efficient manner. For example, if $\calA_{\lambda_\ell} = \calA_{\lambda_k}$, \code{glmnet} does not fit the OLS model for $\lambda_\ell$ but sets $\hat\beta^{LS}(\lambda_\ell) = \hat\beta^{LS}(\lambda_k)$.

The returned object has a \code{predict} method which the user can use to make predictions on future data. As an example, the code below returns the relaxed lasso predictions for the training data at $\gamma = 0.5$ (the default value is \code{gamma = 1}, i.e. the lasso estimator):
\begin{CodeChunk}
\begin{CodeInput}
R> predict(fit, x, gamma = 0.5)
\end{CodeInput}
\end{CodeChunk}
The \code{cv.glmnet} function works with the relaxed lasso as well. When cross-validating a relaxed lasso model, \code{cv.glmnet} provides optimal values for both the \code{lambda} and \code{gamma} parameters. We note that we can consider as many values of the $\gamma$ hyperparameter as we like in CV. Most of the computational time is spent obtaining $\hat\beta^{lasso}(\lambda)$ and $\hat\beta^{LS}(\lambda)$; once we have have computed them $\hat\beta^{relax}(\lambda, \gamma)$ is simply a linear combination of the two. By default, \code{cv.glmnet} performs CV for \code{gamma = c(0, 0.25, 0.5, 0.75, 1)}.

In the exposition above we have focused on the \code{family = "gaussian"} case. Relaxed fits are also available for the rest of the other model families, i.e. any other \code{family} argument. Instead of fitting the OLS model of the response on the active set to obtain the relaxed fit, \code{glmnet} fits the unpenalized model for that model family on the active set.

We note that while the relaxation can be applied for $\alpha$ values smaller than 1, we do not recommend doing this. Relaxation is typically applied to obtain sparser models. It achieves this by undoing shrinkage of coefficients in the active set toward zero, allowing the model to have more freedom to fit the response. Together with CV on $\lambda$ and $\gamma$, this often gives us a model that is sparser than the lasso. Selecting $\alpha$ smaller than 1 results in a larger active set than that for the lasso, working against the goal of obtaining a sparser model.

\subsection{Application to forward stepwise regression}

One use case for the relaxed fit is to provide a faster version of forward stepwise regression. When the number of variables $p$ is large, forward stepwise regression can be tedious since it only adds one variable at a time and at each step, it needs to try all predictor variables that are not already included in the model to find the best one to be added. On the other hand, because the lasso solves a convex problem, it can identify good candidate sets of variables over 100 values of the $\lambda$ hyperparameter even when $p$ is in the tens of thousands. In a case like this, one can run \code{cv.glmnet} and fit the OLS model for a sequence of selected variable sets.
\begin{CodeChunk}
\begin{CodeInput}
R> fitr <- cv.glmnet(x, y, gamma = 0, relax = TRUE)
\end{CodeInput}
\end{CodeChunk}
\section{Assessing models}\label{sec:assess}

After fitting elastic net models with \code{glmnet}, we often want to assess their performance on a set of evaluation or test data. After deciding on the performance measure, for each model in the fitted sequence (indexed by the value of $\lambda$ and possibly $\gamma$ for relaxed fits) we have to build a matrix of predictions and compute the performance measure for it.

\code{cv.glmnet} does some of this evaluation automatically. In performing CV, \code{cv.glmnet} computes the pre-validated fits \citep{Tibshirani2002}, that is the model's predictions of the linear predictor on the held-out fold, and then computes the performance measure with these pre-validated fits. The performance measures are recorded in the \code{cvm} element of the returned \code{cv.glmnet} and are used to make the CV plot when the \code{plot} method is called.

\pkg{glmnet} supports a variety of performance measures depending on the model family: the full list of measures can be seen via the call \code{glmnet.measures()}. The user can change the performance measure computed in CV by specifying the \code{type.measure} argument. For example, the code below computes the area under the curve (AUC) of the pre-validated fits instead of the deviance which is the default for \code{family = "binomial"}:
\begin{CodeChunk}
\begin{CodeInput}
R> fitr <- cv.glmnet(x, y, family = "binomial", type.measure = "auc")
\end{CodeInput}
\end{CodeChunk}
More generally, model assessment can be performed using the \code{assess.glmnet} function. The user can pass a matrix of predictions, a class \code{"glmnet"} object, or a class \code{"cv.glmnet"} object to \code{assess.glmnet} along with the true response values. The code below shows how one can use \code{assess.glmnet} with these three objects, where the training design matrix and response is \code{x[itrain, ]} and \code{y[itrain]} respectively and the testing design matrix and response is \code{x[-itrain, ]} and \code{y[-itrain]} respectively.
\begin{CodeChunk}
\begin{CodeInput}
R> fit <- glmnet(x[itrain, ], y[itrain])
R> assess.glmnet(fit, newx = x[-itrain, ], newy = y[-itrain])

R> pred <- predict(fit, newx = x[-itrain, ])
R> assess.glmnet(pred, newy = y[-itrain])

R> cfit <- cv.glmnet(x[itrain, ], y[itrain])
R> assess.glmnet(cfit, newx = x[-itrain, ], newy = y[-itrain])
\end{CodeInput}
\end{CodeChunk}
By default \code{assess.glmnet} will return all possible performance measures for the model family. Note that if a matrix of predictions is passed, the user has to specify the model family via \code{family} argument since \code{assess.glmnet} cannot infer that from the inputs. (The default value for the \code{family} argument is \code{"gaussian"}, which is what would have been used in the code above.) If a class \code{"glmnet"} object is passed to \code{assess.glmnet}, it returns one performance measure value for each model in the $\lambda$ sequence while if a class \code{"cv.glmnet"} object is passed, it returns the performance measure value at the \code{lambda.1se} value of the $\lambda$ hyperparameter. The user can get the performance measure values at other values of the hyperparameters using the \code{s} and \code{gamma} arguments as in the \code{predict} method.

One major use of \code{assess.glmnet} is to avoid running CV multiple times to get the values for different performance measures. By default, \code{cv.glmnet} will only return a single performance measure. However, if the user specifies \code{keep = TRUE} in the \code{cv.glmnet} call, the pre-validated fits are returned as well. The user can then pass the pre-validated matrix to \code{assess.glmnet}. The code below is an example of how to do this for the Poisson model family. (The \code{keep} argument is \code{FALSE} by default as the pre-validated matrix is large when the number of training observations is large, thus inflating the size of the returned object.)
\begin{CodeChunk}
\begin{CodeInput}
R> cfit <- cv.glmnet(x[itrain, ], y[itrain], keep = TRUE)
R> assess.glmnet(cfit$fit.preval, newy = y, family = "poisson")
\end{CodeInput}
\end{CodeChunk}
We have two additional functions that provide test performance which are unique to binomial data. As the function names suggest, \code{roc.glmnet} and \code{confusion.glmnet} produce the receiver operating characteristic (ROC) curve and the confusion matrix respectively for the test data. Here is an example of the output the user gets from \code{confusion.glmnet}:
\begin{CodeChunk}
\begin{CodeInput}
R> data(MultinomialExample)
R> set.seed(101)
R> itrain <- sample(1:500, 400, replace = FALSE)
R> cfit <- cv.glmnet(x[itrain, ], y[itrain], family = "multinomial")
R> cnf <- confusion.glmnet(cfit, newx = x[-itrain, ], newy = y[-itrain])
R> print(cnf)
\end{CodeInput}
\begin{CodeOutput}
         True
Predicted  1  2  3 Total
    1     13  6  4    23
    2      7 25  5    37
    3      4  3 33    40
    Total 24 34 42   100

 Percent Correct:  0.71
\end{CodeOutput}
\end{CodeChunk}
\section{Discussion} \label{sec:summary}

We have shown how to extend the use of the elastic net penalty to all GLM model families, Cox models with (start, stop] data and with strata, and to a simplified version of the relaxed lasso. We have also discussed how users can use the \pkg{glmnet} package to assess the fit of these elastic net models. These new capabilities are available in version 4.1 and later of the \pkg{glmnet} package on CRAN.

\section*{Acknowledgments}

We would like to thank Robert Tibshirani for helpful discussions and comments. Balasubramanian Narasimhan's work is funded by Stanford Clinical \& Translational Science Award grant 5UL1TR003142-02 from the NIH National Center for Advancing Translational Sciences (NCATS). Trevor Hastie was partially supported by grants DMS-2013736 and IIS
1837931 from the National Science Foundation, and grant 5R01 EB
001988-21 from the National Institutes of Health.

\bibliography{refs}

\end{document}